\documentstyle[prl,twocolumn,aps,english]{revtex}

\begin{document}

\title{Variability of Fundamental Constants}

\author{Asher Peres}
\address{Department of Physics, Technion---Israel Institute of
Technology, 32000 Haifa, Israel}

\maketitle
\begin{abstract}
If the fine structure constant is not really constant, is this due to a
variation of $e$, $\hbar$, or $c$? It is argued that the only reasonable
conclusion is a variable speed of light.
\end{abstract}

\bigskip

There have recently been indications that the fine structure constant
slowly changes over cosmological times \cite{webb}.
There is a lively controversy \cite{duff} whether this is due to
variations of $e$, $\hbar$, or $c$. In this note, it is argued that
the speed of light $c$ may not be constant.

It is customary in theoretical texts to use ``natural units''
$c=\hbar=1$. The rationale is that relativistic transformations mix
time and space coordinates, so that $c$ is only a conversion factor
which can conveniently be set to unity. However, relativistic
transformations never change the nature of timelike and spacelike
intervals. Time is {\it not\/} a fourth dimension of space.

Likewise $h$ can be used to convert joules to hertz and it is
convenient to set $h=2\pi$. Yet, energy and frequency are different
concepts and Planck's constant is more than a conversion factor. Quantum
systems are not localized, they are pervasive. In particular, entangled
systems may be spread over arbitrary distances. A value of $\hbar$
varying in spacetime would necessitate a complete revision of quantum
theory.

Can the electric charge vary? It is a historical accident that Coulomb's
law of force between macroscopic charges was discovered before it was
known that the electric charges of all particles are integral multiples
of $e$ (or $e/3$ if we include quarks). This indicates that we should
define $e=1$ as the unit of charge (this is a {\it natural\/} unit,
not a conversion factor).

We must therefore have a closer look at $c$. In general relativity,
all four coordinates may have different dimensions and the constant
$c$ does not appear in the fundamental equations (it may appear in
particular solutions, once sources that are not generally covariant
have been specified with arbitrary units). In the early universe, where
background radiation cannot be ignored, Lorentz invariance does not hold.
There is a preferred frame. It is then plausible that, in such an
environment, the ``vacuum'' behaves as a dielectric medium where the
speed of light acquires a factor $(\epsilon\mu)^{-1/2}$ \cite{ll}. It
thus appears that a variable $c$ is the most reasonable conclusion to
be drawn from a variation of the fine structure constant. However this
phenomenon is not yet fully understood and more work is clearly needed.

\end{document}